\documentclass[%
 aip,
 amsmath,amssymb,
 reprint,%
]{revtex4-1}

\usepackage{graphicx}
\usepackage{dcolumn}
\usepackage{bm}

\usepackage[utf8]{inputenc}
\usepackage[T1]{fontenc}
\usepackage{mathptmx}

\begin{document}

\preprint{AIP/123-QED}

\title[SOTFET: Proposal for a New Magnetoelectric Memory]{Spin-Orbit Torque Field-Effect Transistor (SOTFET):\\ Proposal for a New Magnetoelectric Memory}

\author{Xiang Li}
\email{xl633@cornell.edu}
\affiliation{School of Electrical and Computer Engineering, Cornell University, Ithaca, NY 14853, USA.}
\author{Joseph Casamento}
\affiliation{Department of Materials Science and Engineering, Cornell University, Ithaca, NY 14853, USA.}
\author{Phillip Dang}
\affiliation{School of Applied and Engineering Physics, Cornell University, Ithaca, NY 14853, USA.}
\author{Zexuan Zhang}
\affiliation{School of Electrical and Computer Engineering, Cornell University, Ithaca, NY 14853, USA.}
\author{Olalekan Afuye}
\affiliation{School of Electrical and Computer Engineering, Cornell University, Ithaca, NY 14853, USA.}
\author{Antonio B. Mei}
\affiliation{Department of Materials Science and Engineering, Cornell University, Ithaca, NY 14853, USA.}
\author{Alyssa B. Apsel}
\affiliation{School of Electrical and Computer Engineering, Cornell University, Ithaca, NY 14853, USA.}
\author{Darrell G. Schlom}
\affiliation{Department of Materials Science and Engineering, Cornell University, Ithaca, NY 14853, USA.}
\author{Debdeep Jena}
\affiliation{School of Electrical and Computer Engineering, Cornell University, Ithaca, NY 14853, USA.}
\affiliation{Department of Materials Science and Engineering, Cornell University, Ithaca, NY 14853, USA.}
\affiliation{Kavli Institute at Cornell for Nanoscale Science, Ithaca, NY 14853, USA.}
\author{Daniel C. Ralph}
\affiliation{Kavli Institute at Cornell for Nanoscale Science, Ithaca, NY 14853, USA.}
\affiliation{Department of Physics, Cornell University, Ithaca, NY 14853, USA.}
\author{Huili Grace Xing}
\affiliation{School of Electrical and Computer Engineering, Cornell University, Ithaca, NY 14853, USA.}
\affiliation{Department of Materials Science and Engineering, Cornell University, Ithaca, NY 14853, USA.}
\affiliation{Kavli Institute at Cornell for Nanoscale Science, Ithaca, NY 14853, USA.}

\date{\today}

\begin{abstract}
Spin-based memories are attractive for their non-volatility and high durability but provide modest resistance changes, whereas semiconductor logic transistors are capable of large resistance changes, but lack memory function with high durability. The recent availability of multiferroic materials provides an opportunity to directly couple the change in spin states of a magnetic memory to a charge change in a semiconductor transistor.  In this work, we propose and analyze the spin-orbit torque field-effect transistor (SOTFET), a device with the potential to significantly boost the energy efficiency of spin-based memories, and to simultaneously offer a palette of new functionalities.
\end{abstract}

\maketitle

The understanding of spin transport in heterostructures \cite{Hellman17} led to the realization of magnetic memories based on giant magnetoresistance (GMR) \cite{Baibich88,Fert08,Grunberg08} and spin-transfer torque (STT) \cite{Slonczewski89,Slonczewski96,Berger96,Ralph08}. Current research aims to make the writing process for magnetic memories more efficient using spin-orbit torques (SOTs) \cite{Miron11,LuqiaoSci12}. STT and SOT magnetic random access memories (MRAMs) offer the virtues of non-volatility, infinite endurance, and good write speeds \cite{Ralph08}.  Nonetheless, the modest resistance change between the magnetic `0' and `1' states of STT- and SOT-MRAMs necessitates a substantial current to obtain acceptable readout voltages, impairing read energies and speeds.

In contrast, non-magnetic semiconductor field-effect transistors (FETs) achieve orders of magnitude resistance change in each switching event. The field effect converts a linear change in the gate voltage into an exponential change in the mobile carrier density in the semiconductor, and consequently modulates its resistance. Thus, a material that can transduce the change of the spin/magnetic state in a SOT structure into the charge of a semiconductor channel could significantly boost the resistance change of a magnetic memory.

This requirement would be met by recently developed magnetoelectric multiferroic materials, which simultaneously possess magnetic order and ferroelectricity in a manner that these order parameters are coupled due to the magnetoelectric effect \cite{Ramesh07,Fiebig16,Chu07}. Moreover, exchange coupling of spins in a ferromagnetic layer to the magnetic order of a multiferroic layer across ferromagnet/multiferroic heterointerfaces has been experimentally demonstrated \cite{Heron14,Qiu13}. 

Inspired by these recent advances in SOT and multiferroic materials, we propose a new magnetoelectric memory device, the spin-orbit-torque field-effect transistor (SOTFET). This device aims to combine the virtues of magnetic memories with the large resistance change of FETs, providing both memory and logic functionalities. Analysis of the memory aspect indicates that the SOTFET can offer orders of magnitude increase in the on-off resistance ratio compared to existing magnetic memories, which can potentially lower the operation energy significantly. The potential logic aspect of the SOTFET would also enable new circuit architectures for efficient logic or search functions \cite{Lekan19}. In this paper, we present the physical operation of the SOTFET along with a device model, and will mainly focus on the memory aspect.

\begin{figure}[ht]
\centering
\includegraphics[width=0.85\columnwidth]{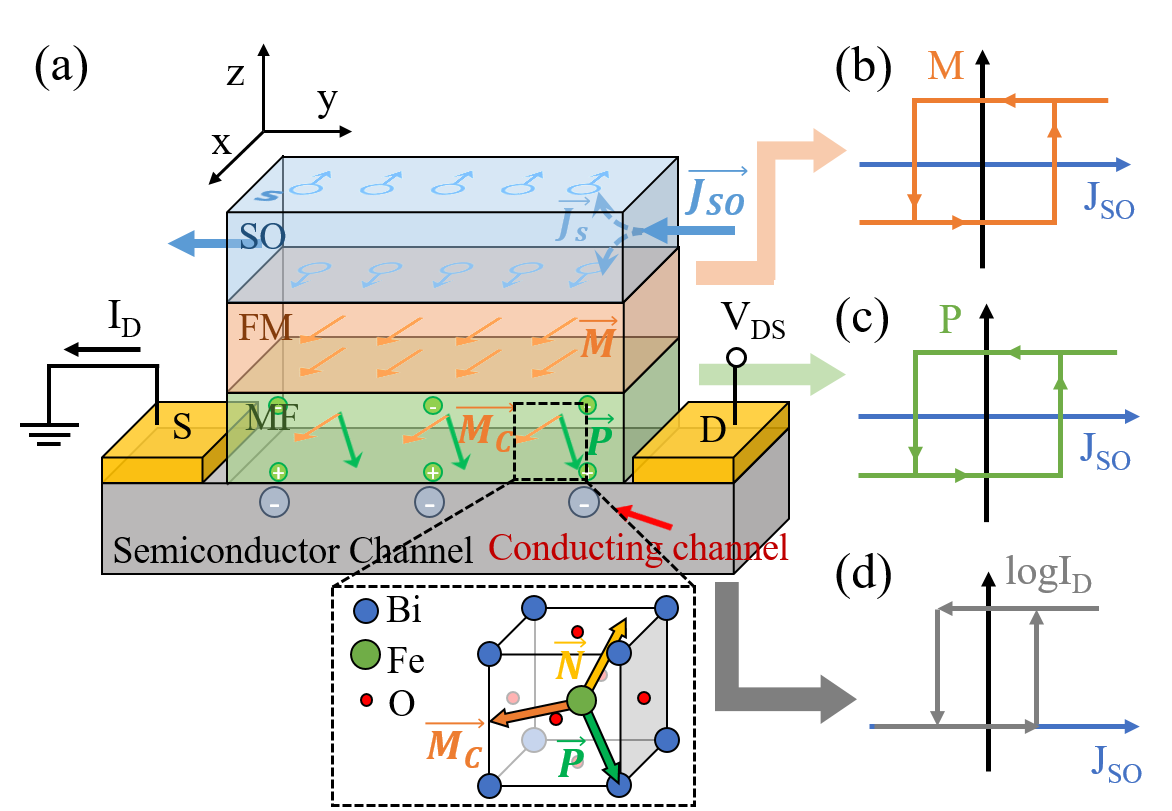}
\caption{\label{fig:structure} (a) Device structure and working principle of a SOTFET. A CoFe/BiFeO$_3$ bilayer is employed in this study as the example FM/MF bilayer. The \textbf{P}, \textbf{M}$_\text{C}$, and \textbf{N} in BiFeO$_3$ are indicated in its perovskite unit cell. In equilibrium, \textbf{P} points to one of the $\langle$111$\rangle$ directions. (b) A charge current $J_{SO}$ through the spin-orbit (SO) layer switches the magnetization M in the FM layer, (c) which in turn switches the polarization P in the MF layer.  As a result, (d) the semiconductor channel resistance is modulated and the drain current $I_D$ is used as the read-out component.}
\end{figure}

Figure~\ref{fig:structure} shows the structure of a SOTFET. It resembles an ordinary metal-oxide-semiconductor FET (MOSFET), but with a unique gate stack. The SOTFET gate stack comprises three layers (from top to bottom): a spin-orbit (SO) layer, a ferromagnetic (FM) layer, and a multiferroic (MF) layer, adjacent to a semiconductor channel to which source and drain contacts are made. 

The working principle of the SOTFET is illustrated in Fig.~\ref{fig:structure}. The state of magnetization \textbf{M} of the FM layer is the memory component. When a charge current $J_{SO}$ flows in the SO layer, transverse spin-polarized currents are generated due to spin-momentum locking \cite{Dyakonov71,Hirsch99,Sinova15,Bychkov84,Fu07,Hasan10}. Spin absorption at the SO/FM interface exerts a spin-orbit torque that switches the \textbf{M} of the FM \cite{Ralph08,Mellnik14,Luqiao12}, as illustrated in Fig.~\ref{fig:structure}(a) and qualitatively plotted in Fig.~\ref{fig:structure}(b). Flowing $J_{SO}$ in the opposite direction switches the magnetization between `1' $\leftrightarrow$ `0', identical to the conventional writing mechanism in SOT-MRAMs.

The SOTFET differs from the conventional SOT-MRAMs in the read mechanism. Coupling the \textbf{M} of the FM with the semiconductor channel would be achieved by the magnetoelectric multiferroic layer. Due to the exchange coupling between the FM and the MF \cite{Heron14,Qiu13}, the magnetic dipole of the MF is also switched with the \textbf{M} in the FM. Within the MF material, the Dzyaloshinskii-Moriya interaction (DMI) \cite{Dzyaloshinsky58,Moriya60} effectively couples electric and magnetic dipoles, since the weak canted magnetic moment $\textbf{M}_{\text{C}}$ originates from the DMI \cite{Heron11,Heron14,Ederer05}. When the $\textbf{M}_{\text{C}}$ switches polarity, the electric polarization ${\bf P}$ in the MF deterministically switches in tandem, all in response to $J_{SO}$, as indicated in Fig.~\ref{fig:structure}(c).

The resulting switching of \textbf{P} gates the semiconductor channel by shifting the surface potential, similar to the effect in ferroelectric-gate FETs \cite{George16,Scott89,Miller92,Salahuddin08}. The current $I_{D}$ flowing in the semiconductor channel is the read-out signal, which changes by several orders of magnitude due to the resistance change. Consider the direction of $J_{SO}$ in the SO layer in Fig.~\ref{fig:structure}(a) as writing a `1' in the FM, leading to a high conductivity ON state of the semiconductor. When the current $J_{SO}$ flows in the opposite direction, all the dipoles in the gate stack are flipped. The flipping of ${\bf P}$ then depletes the semiconductor channel, putting it in the OFF state.  The resulting transistor output current $I_D$ in response to $J_{SO}$ is shown in Fig.~\ref{fig:structure}(d): it is bi-stable, and provides the desired large resistance ratio for efficient readout.

To prove the feasibility of the SOTFET, we quantitatively analyze the dynamical coupling across each interface, and the entire device. The analysis to follow shows that the SOTFET behavior is achievable, but requires magnetoelectric multiferroics of specific magnetism and polarization, along with an appropriate hierarchy of strengths for the exchange coupling, DMI, and anisotropy energies within the gate stack.  Because the FM/MF heterointerface is the least explored, instead of a generic case, we use experimental results of the CoFe/BiFeO$_3$ heterostructure \cite{Heron14,Heron11} in this initial exploration. Bi$_2$Se$_3$ is selected as the example SO layer. Other materials candidates are discussed in \cite{Phillip19}. The aim of the model is to guide experiments by pointing towards desired heterointerface choices.

The magnetization \textbf{M} of the FM layer is switched by spin-orbit torque (SOT).  For simplicity we assume single-domain macrospin behavior. The switching dynamics of this process are captured by the Landau-Lifshitz-Gilbert-Slonczewski (LLGS) equation \cite{Slonczewski96,Ralph08,Xiao05,Bazaliy13}:
\begin{equation}
    \label{eq:LLGS}
    \frac{d\hat{\textbf{m}}}{dt}
    =-\gamma\mu_0\hat{\textbf{m}}\times\textbf{H}_{eff}
    +\alpha\hat{\textbf{m}}\times\frac{d\hat{\textbf{m}}}{dt}
    +(\frac{\gamma}{M_S})\vec{\tau}_{SOT},
\end{equation}
where $\hat{\textbf{m}}$ is the normalized magnetization of the FM, $\textbf{H}_{eff}$ is the effective magnetic field acting on $\hat{\textbf{m}}$, $\gamma$ is the electron gyromagnetic ratio, $\mu_0$ is the vacuum permeability, $\alpha$ is the Gilbert damping factor, $M_S$ is the saturation magnetization and $\vec{\tau}_{SOT}=\vec{\tau}_{AD}+\vec{\tau}_{FL}$ is the spin-orbit torque, the sum of the anti-damping torque $\vec{\tau}_{AD}$ and field-like torque $\vec{\tau}_{FL}$, which are given by \cite{Ralph08,Xiao05}:
\begin{eqnarray}
    \label{eq:SOT AD torque}
    &\vec{\tau}_{AD}=(\frac{\hbar}{2e})(\frac{1}{t})j
    \theta_{AD}\hat{\textbf{m}}\times(\hat{\textbf{m}}\times\hat{\textbf{m}}_p), \text{ and } \\
    \label{eq:SOT FL torque}
    &\vec{\tau}_{FL}=(\frac{\hbar}{2e})(\frac{1}{t})j\theta_{FL}\hat{\textbf{m}}\times\hat{\textbf{m}}_p.
\end{eqnarray}
Here $\hbar$ is the reduced Planck constant, $e$ is electron charge, $t$ is the thickness of ferromagnetic (FM) material, $j=J_{SO}$ is the charge current density in the SO layer, $\theta_{AD(FL)}$ is the spin Hall angle of the anti-damping (AD) or field-like (FL) torque from the SO layer, and $\hat{\textbf{m}}_p$ is the normalized spin polarization.

The effective field $\textbf{H}_{eff}=\textbf{H}_{ext}+\textbf{H}_a+\textbf{H}_{demag}+\textbf{H}_{DMI}$, where $\textbf{H}_{ext}$ is any external magnetic field and $\textbf{H}_a$ is the anisotropy field with perpendicular magnetic anisotropy (PMA) calculated by $\textbf{H}_a=\frac{2K}{\mu_0M_S}m_z\hat{z}\equiv H_km_z\hat{z}$ \cite{Bazaliy13}, where $K$ is the anisotropy constant. $\textbf{H}_{demag}$ is the demagnetization field as calculated in Beleggia et al.~\cite{Beleggia06}. The last term $\textbf{H}_{DMI}$ is the effective magnetic field arising from the effective DMI, which is discussed further below.

Switching of \textbf{M} in the FM switches the electric polarization \textbf{P} of the MF due to the exchange coupling and DMI. The dynamics of \textbf{P} are captured by the Landau-Khalatnikov (LK) equation \cite{Landau54,Salahuddin08,Liao19}:
\begin{equation}
    \label{eq:LK}
    \gamma_{FE}\frac{\partial P_i}{\partial t}=-\frac{\partial F}{\partial P_i},
\end{equation}
where $\gamma_{FE}$ is the viscosity coefficient, $P_i(i=x, y, z)$ is the $x/y/z$ component of \textbf{P}. $F$ is the total ferroelectric free energy \cite{Zhang07,Liao19}:
\begin{eqnarray}
    \label{eq:free energy}
    & F(\textbf{P},\textbf{u})
    =\alpha_1(P_x^2+P_y^2+P_z^2)
    +\alpha_{11}(P_x^4+P_y^4+P_z^4)
    \nonumber\\
    & +\alpha_{12}(P_x^2P_y^2+P_x^2P_z^2+P_y^2P_z^2)
    +K_{strain}(\hat{\textbf{P}}\cdotp\textbf{u})^2
    \nonumber\\
    & -\textbf{P}\cdotp(\textbf{F}_{ext}+\textbf{F}_{DMI}),
\end{eqnarray}
where $\alpha_1$, $\alpha_{11}$, $\alpha_{12}$ are the phenomenological Landau expansion coefficients, $K_{strain}$ is the strain energy, \textbf{u} is the axis of substrate strain, \textbf{F}$_{ext}$ is the external electric field and \textbf{F}$_{DMI}$ is the effective electric field from DMI. The strain term, $K_{strain}(\textbf{P}\cdotp\textbf{u})^2$, in Eq.~\ref{eq:free energy} arises from the substrate-induced strain \cite{Liao19,Zhang07}, which dictates the energy-favorable planes for the equilibrium states of $\textbf{P}$, thereby reducing the degeneracy of $\textbf{P}$ orientations in the specific case of the MF BiFeO$_3$, which is shown in previous studies \cite{Heron14,Li04,Mei19}.

In our model, the exchange coupling, which couples \textbf{M}$_{\text C}$ in BiFeO$_3$ and \textbf{M} in CoFe, and the DMI, which couples \textbf{P} and \textbf{M}$_{\text C}$ in BiFeO$_3$, are merged into one effective DMI term that directly captures the interaction between the \textbf{M} in CoFe and $\textbf{P}$ in BiFeO$_3$, with an effective Hamiltonian \cite{Ederer05,Heron14}:
\begin{equation}
    \label{eq:DMI_Hamiltonian}
    E_{DMI}=-E_{DMI,0}\hat{\textbf{P}}\cdotp (\hat{\textbf{N}}\times\hat{\textbf{M}}),
\end{equation}
where $E_{DMI,0}$ is the energy coefficient of DMI, $\hat{\textbf{P}}$ is the polarization of BiFeO$_3$, $\hat{\textbf{N}}$ is the Neel vector, and $\hat{\textbf{M}}$ is the magnetic moment in CoFe. All vectors in the equation are normalized vectors. The effective magnetic and electric fields that enter the equations of motions are then:
\begin{eqnarray}
    \label{eq:H_DMI}
    \textbf{H}_{DMI}=-\frac{1}{\mu_0M_S}\frac{\partial E_{DMI}}{\partial\hat{\textbf{M}}}\equiv H_{DMI,0}(\hat{\textbf{P}}\times\hat{\textbf{N}})\\
    \label{eq:F_DMI}
    \text{ and }\textbf{F}_{DMI}=-\frac{1}{P_S}\frac{\partial E_{DMI}}{\partial\hat{\textbf{P}}}\equiv F_{DMI,0}(\hat{\textbf{N}}\times\hat{\textbf{M}}),
\end{eqnarray}
where $H_{DMI,0}$ is the effective DMI magnetic field magnitude and $F_{DMI,0}$ is the effective DMI electric field magnitude. Both fields have constant magnitudes for specific material combinations, because they originate from the energy and material parameters:
\begin{equation}
    \label{eq:DMI_energy}
    E_{DMI,0}=\mu_0M_S\cdotp H_{DMI,0}=P_S\cdotp F_{DMI,0}.
\end{equation}

\begin{figure}[ht]
\includegraphics[width=0.8\columnwidth]{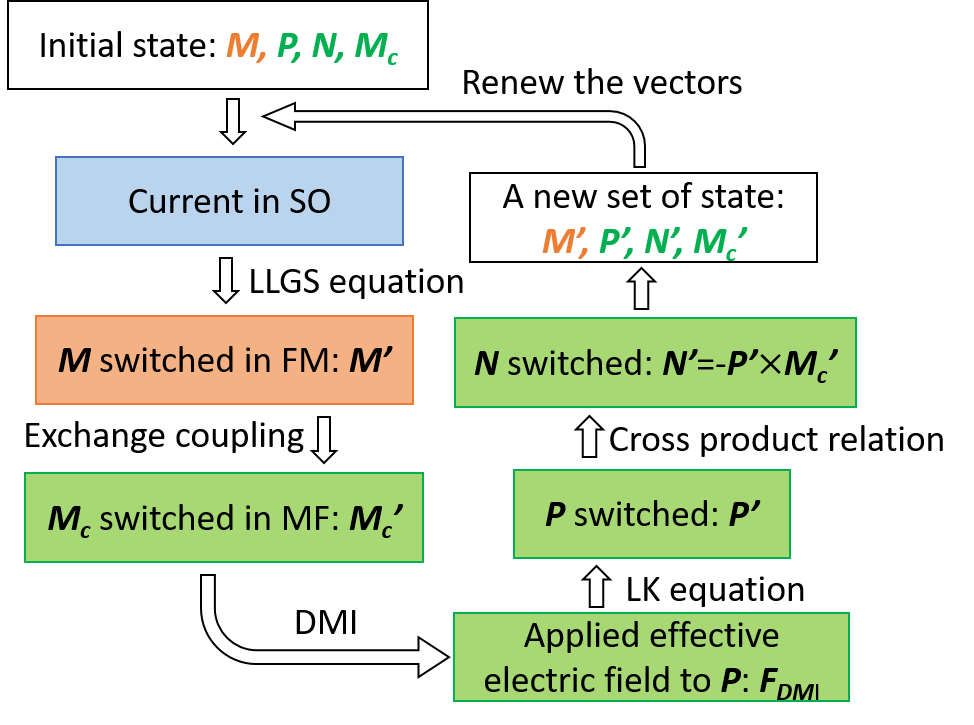}
\caption{\label{fig:procedure} Modeling procedure flow of the SOTFET model.}
\end{figure}

With the direction of \textbf{N} defined as $\hat{\textbf{N}}=-\hat{\textbf{P}}\times\hat{\textbf{M}}$, all vectors in the CoFe/BiFeO$_3$ FM/MF system ($\textbf{P}$, $\textbf{N}$ and $\textbf{M}$) are connected by the DMI. The method of implementing the dynamic evolution of the \textbf{M} and \textbf{P} described above is shown schematically in Fig.~\ref{fig:procedure}.  The initial state of the SOTFET is defined by a set of vectors: \textbf{M} in the FM, and $\textbf{M}_{\text C}$, \textbf{P}, and \textbf{N} in the MF. When a current $J_{SO}$ flows in the SO layer, all 4 vectors (\textbf{M}, $\textbf{M}_{\text C}$, \textbf{P}, and \textbf{N}) can switch to new states, with dynamics dictated by the LLGS and the LK equations in each loop. Finally, a new set of the 4 vectors in a new equilibrium state will be reached by iteration. The switching behavior of \textbf{P} and \textbf{M} is assumed to be purely rotational, consistent with experimental studies of BiFeO$_3$ \cite{Heron14}.

\begin{figure*}
\centering
    \includegraphics[width=1.97\columnwidth]{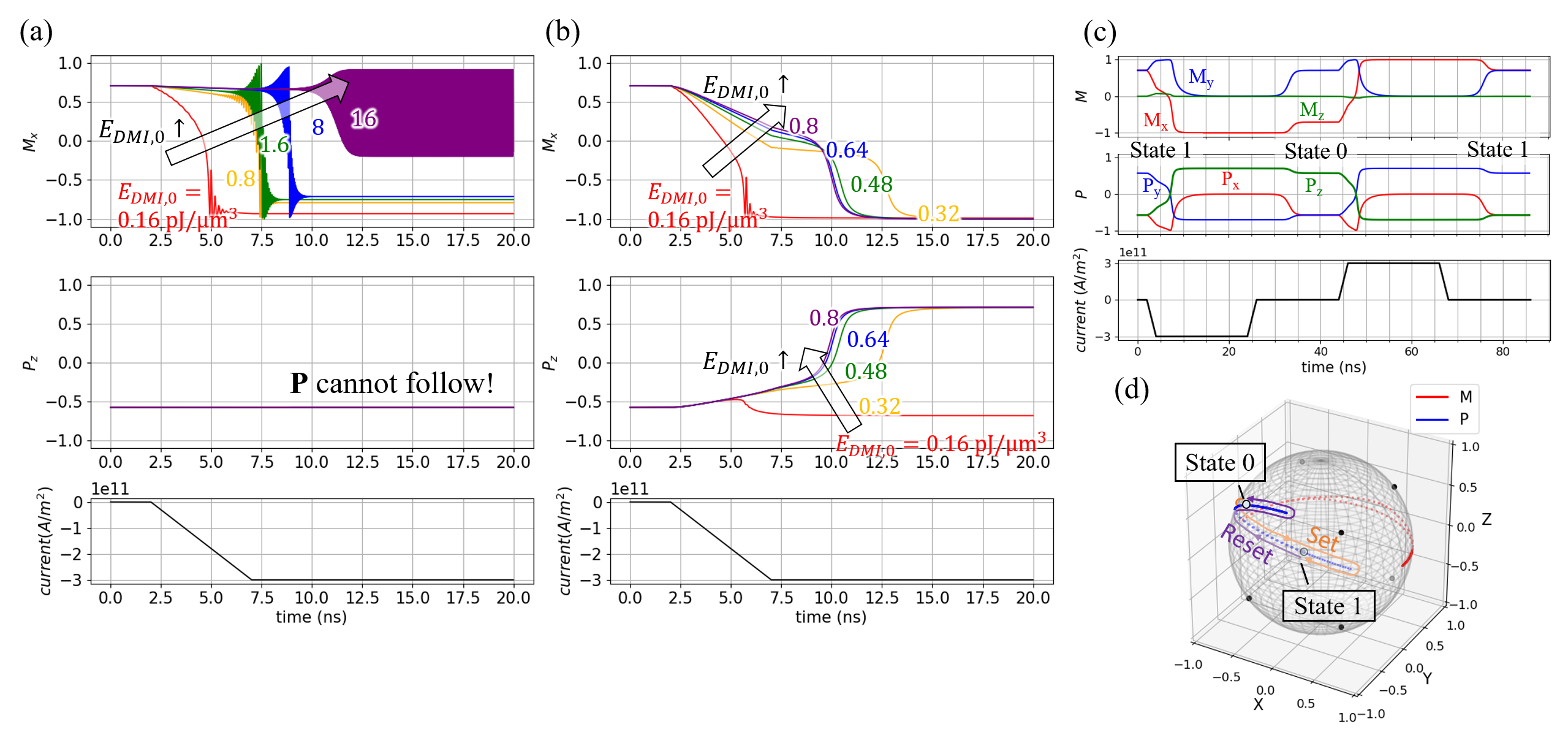}
    \caption{Switching behavior in a SOTFET gate stack for a range of DMI energy assuming (a) $P_S$=100 $\mu$C/cm$^2$, and (b) $P_S$=10 $\mu$C/cm$^2$. For a high $P_S$ in (a), it is observed that upon applying a current $J_{SO}$ (lower panel), $M_x$ responds to the spin-orbit torque (upper panel); $P_z$, however, does not switch (mid panel). For a lower $P_S$ in (b), it is observed that above a critical DMI energy, both $M_x$ and $P_z$ switch deterministically. (c) and (d) \textbf{M} and \textbf{P} can be switched by $J_{SO}$ into “0” and “1” states while showing non-volatility; $P_S$=10 $\mu$C/cm$^2$ and $E_{DMI,0}$=0.8 pJ/$\mu$m$^3$ are assumed. (d) shows the trajectories of \textbf{M} (red) and \textbf{P} (blue) on a sphere. The set to “1” (\textbf{P}=[-1,1,-1]) process is marked by orange and the reset to “0” (\textbf{P}=[-1,-1,1]) by purple.}
    \label{fig:combined results}
\end{figure*}

Key parameters used in the numerical evaluation of the SOTFET are provided in Supplement S1.  The model is validated by comparing to the micromagnetic simulation tools OOMMF \cite{OOMMF} and MuMax3 \cite{MuMax3}, and other theoretical calculations and experimental results, shown in Supplement S2. For the SOTFET gate stack to controllably gate the semiconductor channel, a deterministic switching of the polarization ${\bf P}$ in $z$-direction in the MF is desired. 

For the CoFe/BiFeO$_3$ FM/MF heterostructure and taking $P_S$=100 $\mu$C/cm$^2$ of BiFeO$_3$ \cite{Wang03} and $M_S$=1.6$\times$10$^6$ A/m of CoFe \cite{Qiu13}, switching behavior for a range of DMI energies is shown in Fig.~\ref{fig:combined results}(a). Upon applying a current $J_{SO} = -30 $ MA/cm$^2$, different switching behavior of the $x-$component of the magnetization ($M_x$) is observed for different DMI energies. For these values, however, the $z-$component of the polarization $P_z$ in the MF layer does {\em not} follow the motions of \textbf{M}. This is because, given the large $P_{S}$, a moderate DMI energy is not sufficient to overcome the anisotropy energy in \textbf{P} to switch it.  For a high DMI energy, with \textbf{P} held in place, \textbf{M} also does not switch because the $H_{DMI}$ then functions as an effective unidirectional anisotropy acting back on \textbf{M}. This is therefore a situation when the SOTFET does not achieve the desired functionality.

The natural next step is to explore reduced $P_S$ in the MF layer. Reducing $P_S$ in BiFeO$_3$ is experimentally feasible, for example, by La-substitution of Bi in BiFeO$_3$ \cite{Singl06,Vazquez12}. Recent experiments by Lin et al \cite{Lin19} also shows that the exchange interaction between CoFe and La-substituted BiFeO$_3$ remains strong even with reduced $P_S$. Qualitatively, this implies that the multiferroic layer should have a relatively weak ferroelectricity, a strong magnetization, and strong coupling between the two order parameters. The calculated results with a reduced $P_S$=10 $\mu$C/cm$^2$ and other parameters unchanged are shown in Fig.~\ref{fig:combined results}(b) for a range of $E_{DMI,0}$. For the same current $J_{SO}$, a critical $E_{DMI,0}$ is observed. Above the critical $E_{DMI,0}$, $M_x$ and $P_z$ concomitantly switch, signalling the required materials parameters for successful SOTFET operation. 

Reducing $P_S$ of the BiFeO$_3$ could help the switching of \textbf{P} successfully for two reasons. First, as shown in Eq.~\ref{eq:DMI_energy}, for a fixed $H_{DMI,0}$ and $M_S$, lowering $P_S$ for the same $E_{DMI,0}$ implies an enhanced $F_{DMI,0}$ to switch the polarization. Second, a reduced $P_S$ leads to a weaker polarization anisotropy as described in the free energy equation Eq.~\ref{eq:free energy}. This lowers the energy barrier between polarization equilibrium states, making the switching easier.

Figures~\ref{fig:combined results}(c) and \ref{fig:combined results}(d) show that the desired stable switching behavior of the SOTFET is achieved by choosing the CoFe/BiFeO$_3$ heterostructure with a reduced $P_S$=10 $\mu$C/cm$^2$ of the MF (BiFeO$_3$) and an above-critical $E_{DMI,0}$=0.8 pJ/$\mu$m$^3$, which corresponds to DMI fields of $H_{DMI,0}$=5 kOe and $F_{DMI,0}$=80 kV/cm. It is seen that switching the current direction in the SO layer successfully switches $P_z$. The current density used, 30 MA/cm$^2$, is about 1 order of magnitude lower than heavy-metal based SOT-MRAMs \cite{Meng16,MengPRB16} due to the assumed large spin Hall angle of Bi$_2$Se$_3$ ($\theta_{AD}=\theta_{FL}=3.5$ \cite{Mellnik14}), and can be further reduced by using larger spin Hall angle materials such as BiSb \cite{Khang18}. \textbf{M} is observed to switch within the $x-y$ plane and \textbf{P} is switched out-of-plane. The switching trajectories of ${\bf M}$ and ${\bf P}$ are shown in the spherical plot in Fig.~\ref{fig:combined results}(d). Clear set and reset processes between State 0 and 1 are observed, proving feasibility of the SOTFET operation for the chosen material parameters.

The switching of \textbf{P} with a reduced $P_S$=10 $\mu$C/cm$^2$ shown in Figs.~\ref{fig:combined results}(c) and \ref{fig:combined results}(d) results in a charge difference $\Delta Q=2P_z\approx$12 $\mu$C/cm$^2$ in the semiconductor channel, assuming the absence of traps at the interface between the MF and the semiconductor channel. For example, for the choice of a silicon channel, this will lead to a surface potential change approximately at $\Delta\psi\approx$1.3 V by a simple calculation \cite{Ytterdal03}, accessing the entire operating regime of a MOSFET from strong inversion to accumulation. Thus, by estimation, at least an on/off ratio of 10$^8$ in $I_D$ can be achieved due to the resistance change of the channel, which in practice will be limited by gate leakage and interfacial trap states rather than the intrinsic capability of a SOTFET. The simulation integrated with a Si-MOSFET model in SPICE verifies that an on/off ratio $>10^7$ can be achieved \cite{Lekan19}, which indicates that a $P_S\approx0.1 \mu C/cm^2$ is sufficient to fully control the semiconductor channel for the high on/off ratio when assuming no defect presents. The high on/off ratio in $I_D$ as the read-out component brings the read energy of a SOTFET down to the same level as a conventional semiconductor transistor.

An electrically insulating magnetic (FM) layer is more desirable for SOTFET application in order to reduce the shunting current from SO layer, and boost the spin torque efficiency \cite{Wang16}. Besides, the insulating FM layer should potentially reduce the charge injection in the MF layer, thus alleviating the fatigue that is often confronted by ferroelectric materials. The fatigue issue is also addressed by the fact that the polarization switching is driven by coupling to the magnetic layer rather than an external electric field, which should reduce the tendency for long-distance atom motion.

Another challenge in the development of ferroelectric memory devices has been the presence of the depolarization field that can destabilize the ferroelectric polarization over time \cite{Jacobs73}.  If the hierarchy of coupling energies within the SOTFET is designed correctly, P cannot switch unless M is switched by $J_{SO}$.  Therefore, the exchange coupling of the multiferroic to the magnetic layer can help to improve the ferroelectric retention.

In addition, by the virtue of simultaneously being a FET, the SOTFET can also provide logic functionality by a gate voltage controlling the channel. As a merger of memory and logic, the SOTFET is capable of process-in-memory (PiM) functionalities that significantly lower the energy consumption and physical size of computation, comparing to a von Neumann architecture where logic and memory are separated. Some examples are explored in \cite{Lekan19}. The experimental realization, and various modes of operation of the SOTFET are currently being investigated.

In summary, the SOTFET, a new magnetoelectric memory device is proposed, in which a change in magnetization of a SO/FM layer is transduced to control the semiconductor channel by using a magnetoelectric multiferroic layer, so that a readout with several orders of magnitude change in resistance can be achieved. We establish a quantitative model of the dynamics of the magnetization and polarization of the layers of the SOTFET. From the model, the materials needs for the successful operation are identified and the feasibility of the SOTFET is proved in a properly designed CoFe/BiFeO$_3$ gate stack.

\textbf{\textit{Supplementary Material -}} Key parameters for numerical simulations and the validation of the model are provided in supplementary material.

\textbf{\textit{Acknowledgments -}}
This work was supported in part by the Semiconductor Research Corporation (SRC) as nCORE task 2758.001 and NSF under the E2CDA program (ECCS 1740286). The authors wish to thank Prof.~Alyosha Molnar, Prof.~Christopher Batten, Yu-Ching Liao, Hyunjea Lee and Yongjian (Helffor) Tang for helpful discussions.

\nocite{*}
\bibliography{aipsamp}

\end{document}